\documentstyle[twocolumn,aps,pra,psfig,floats]{revtex}

\newcommand{\ket}[1]{\left|#1\right\rangle}
\newcommand{\bra}[1]{\langle #1|}
\newcommand{\ketbra}[2]{|#1\rangle\langle #2|}
\newcommand{\braket}[2]{\langle#1|#2 \rangle}
\newcommand{\QKD}{{\sc qkd}}

\newcommand{\USD}{{\sc usd}}

\begin{document}
\draft

\title{Unambiguous state discrimination in quantum cryptography
with weak coherent states}

\author{Miloslav Du\v sek$^1$, Mika Jahma$^2$, and Norbert L\"utkenhaus$^2$}

\address{$^1$Department of Optics, Palack\'y University, 17.~listopadu 50,
         772~00 Olomouc, Czech~Republic}

\address{$^2$Helsinki Institute of Physics, P.O.~Box~9, 00014 Helsingin
         yliopisto, Finland}

\date{\today}

\maketitle


\begin{abstract}
The use of linearly independent signal states in realistic
implementations of quantum key distribution (\QKD) enables an
eavesdropper to perform unambiguous state discrimination. We explore
quantitatively the limits for secure \QKD\ imposed by this fact taking
into account that the receiver can monitor to some extend the photon
number statistics of the signals even with todays standard detection
schemes. We compare our attack to the beamsplitting attack and show that
security against beamsplitting attack does not necessarily imply
security against the attack considered here.
\end{abstract}



\narrowtext

\section{Introduction}

Quantum key distribution (\QKD) is a technique to provide two parties
with a secure, secret and shared key. Such a key is the necessary
ingredient to the only {\em provably\/} secure way to communicate with
guaranteed privacy, the one-time pad or Vernam cipher \cite{vernam26a}.
The first complete protocol was given by Bennett and Brassard
\cite{bennett84a} (BB84) following first ideas by Wiesner
\cite{wiesner83a}. It uses the fact that any channel which transmits two
non-orthogonal states perfectly, automatically makes eavesdropping on
this channel detectable.

We consider the BB84 protocol in a typical quantum optical
implementation. Ideally, Alice sends a sequence of single photons which
are at random polarized in one of the following four states: right or
left circular polarized, or vertically or horizontally polarized. Bob
chooses at random between two polarization analyzers, one distinguishing
the circular polarized states, and the other distinguishing the linear
polarized states. Following a public discussion about the basis of the
sent signals and the measurement apparatus applied to them, sender and
receiver can obtain a shared key made up from those signals where the
measurement device gives deterministic results. This is the {\em sifted
key} \cite{huttner94a}. Proofs of security of this scheme against the
most general attack, even in the presence of noise, have been obtained
\cite{mayers96a,mayers98a,lo99a}. In this article we follow another
goal: we would like to illuminate to which extend already very simple
attacks can render \QKD\ impossible once realistic imperfections like
lossy lines and non-ideal signal states are taken into account. The
difficulties implied, for example, by the use of weak coherent states in
combination with lossy lines has been pointed out earlier
\cite{huttner95a,yuen96a,dusek99a} and this subject has been illuminated
in depth in \cite{brassardsub99a}, where bounds on coverable distances
are given. Positive security proofs for sufficiently short distances and
taking into account the realistic signals are given for individual
attacks \cite{nlsub00a} and coherent attacks \cite{inamorisub00a}. The
eavesdropping attacks which crack the secrecy of the key for set-ups
exceeding this secure distances are still quite complicated. The
eavesdropper needs to perform a QND measurement on the total photon
number of the signal, then he has to split a photon off the occurring
multi-photon signals \cite{brassardsub99a}, store that photon, and then,
finally, measure it after the public discussion.

In this paper we are looking into much simpler eavesdropping strategies
which make use of the  opportunities arising from lossy lines and
non-ideal signals. Such an attack has been proposed by Bennett
\cite{bennett92a} and Yuen \cite{yuen96a}. It uses the fact that Eve
can, with finite probability, discriminate the four signal states
unambiguously. Whenever such a discrimination is performed successfully,
the eavesdropper knows immediately which of the four signal states was
sent and can send this information via a classical channel to Bob's
detector, in front of which she places a state preparation machine to
prepare the identified state. This way this state does not experience
the losses of the actual quantum channel without that Eve has to invest
into a perfect quantum channel.

The investigation of this scenario refines Bennett's and Yuen's analysis
since it takes into account that, to a certain extend, the photon
statistics of the signals arriving at Bob's detectors can be monitored.
The results  illuminate the restrictions placed on implementations of
\QKD\ on lines with strong losses. Thereby we can show that the
currently widely used conditional security standard of security against
beamsplitting attacks \cite{bennett92a} is incomplete. Especially,
contrary to common belief, the use of unambiguous state discrimination
can be a more efficient eavesdropping strategy than the beamsplitting
attack, even for dim coherent states.

The paper is organized as follows. In Sec.~\ref{SECunambiguous} we
recapitulate the principles of unambiguous state discrimination. These
are applied in Sec.~\ref{SECsignal} to the signal states in the BB84
protocol with dim coherent signal states. In Sec.~\ref{SECstrategy} we
introduce an eavesdropping attack based on unambiguous state
discrimination (\USD\ attack) and analyze it in detail, taking the photon
number distribution of the signals arriving at Bob's detectors into
account. In Sec.~\ref{SECUSD} we discuss the relation between the beam
splitting attack and the \USD\  attack. Sec.~\ref{SECconclusions} concludes
the article with a short summary.

\section{Unambiguous discrimination of signal states}
\label{SECunambiguous}

Unambiguous state discrimination is possible whenever the $N$ states in
question are linearly independent. The problem can be described by a
measurement which can give the results ``state 0'', ``state 1'', \dots
``state $N-1$'', and the result ``don't know''. The constraint is that
the measurement results should never wrongly identify a state, and the
goal is to keep the fraction of ``don't know'' results as low as
possible. This problem has been investigated by Ivanovic
\cite{ivanovic87a} for the case of two equal probable non-orthogonal
states. Peres \cite{peres88a} solved this problem in a formulation with
Probability Operator Measures (POM). Later Jaeger and Shimony
\cite{jaeger95a} extended the solution to arbitrary a priori
probabilities. Peres's solution has been
generalized to an arbitrary number of equally probable states which are
generated from each other by a symmetry operator by Chefles and Barnett
\cite{chefles98a}. Their result can be summarized as follows: the
symmetry allows to write the input states in the form
\begin{equation}
\label{standardform}
   \ket{\Psi_k} = \sum_{j=0}^{N-1} c_j \exp \left(2 \pi {\rm i} \frac{k\;
   j}{N}\right) \ket{\phi_j},
\end{equation}
where the states $\ket{\phi_j}$ represent some set of orthonormal
states. Note that the states
 $\ket{\tilde{\Psi}_\ell} =
   \frac{1}{\sqrt{N}}\sum_{j=0}^{N-1}
   \exp\left(2 \pi {\rm i} \frac{\ell\; j}{N} \right) \ket{\phi_j}$
form another orthonormal set. It turns out that the optimal strategy for
unambiguous state discrimination consists of two steps. In the first
step a filter operation is performed such that the output states are
either the orthonormal states $\ket{\tilde{\Psi}_\ell}$ or some linear
dependent states. This step can be described by a complete positive map
with the two Kraus operators. They are  defined with the help of the
minimum coefficient $c_{\rm min} = \min_j |c_j|$ as
\begin{eqnarray}
A_{\rm yes} & = & \sum_{j=0}^{N-1} \frac{c_{\rm min}}{c_j}
\ket{\phi_j}\bra{\phi_j},\\
A_{\rm no} & = & \sum_{j=0}^{N-1} \sqrt{1-\frac{c_{\rm min}^2}{|c_j|^2}}
\ket{\phi_j}\bra{\phi_j}.
\end{eqnarray}
The conditional state in the event  of  successful filtering
is now  given as
$$
\ket{\Psi_k^{(\rm yes)}} = \sqrt{N}c_{\rm min}\ket{\tilde{\Psi}_k}  .
$$

In a second step, we can perform a von Neumann projection measurement on
this state to identify unambiguously the state $k$ via the orthonormal
state $\ket{\tilde{\Psi}_k}$. The probability of this successful
identification is given by
\begin{equation}
P_D = N \min_{j} |c_j|^2  .
\end{equation}
For the case of two equal probable non-orthogonal polarization states of
a single photon a quantum optical implementation following this two-step
idea has been given by Huttner et al.~\cite{huttner96a} and by Brandt
\cite{brandt99a}.

\section{Signal states}
\label{SECsignal}

\begin{figure}[htb]
\centerline{\psfig{width=8cm,file=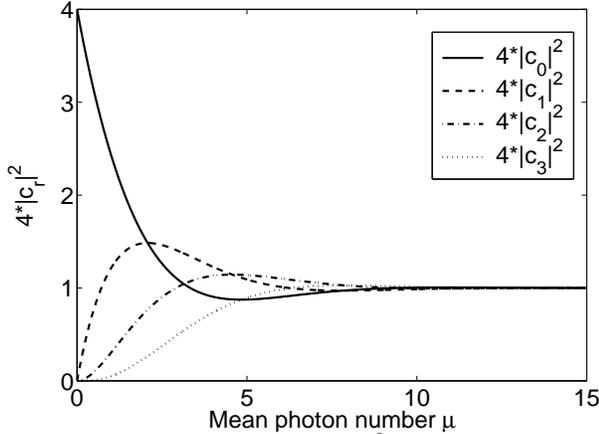}}
\caption{The fourfold weight $4 |c_j|^2$ of the four canonical states
$\ket{\phi_j}$ as a function of the mean photon number $\mu$.
The lower bound of these four curves gives the optimum
probability for unambiguous state discrimination.}
\label{f-1}
\end{figure}

A first description of realistic signal states is that of a coherent
state with a small amplitude $\alpha$. This corresponds to the
description of a dimmed laser pulse. The coherent state is given by
\begin{equation}
\ket{\alpha} = e^{-|\alpha|^2/2} \sum_{n=0}^\infty
\frac{(\alpha a^\dagger)^n}{n!} \ket{0}
\end{equation}
where $a^\dagger$ is the creation operator for one of the four BB84
 polarizations which can be expressed in terms of two creation
 operators $b^\dagger_1$ and $b^\dagger_2$ (corresponding, e.g., to two
 linear orthogonal polarizations) as
\begin{eqnarray}
a^\dagger_0 & = & \frac{1}{\sqrt{2}}\left( b^\dagger_1+b^\dagger_2
\right),\\
a^\dagger_1 & = & \frac{1}{\sqrt{2}}\left( b^\dagger_1+{\rm i}\; b^\dagger_2
\right),\\
a^\dagger_2 & = & \frac{1}{\sqrt{2}}\left( b^\dagger_1- b^\dagger_2
\right),\\
a^\dagger_3 & = & \frac{1}{\sqrt{2}}\left( b^\dagger_1- {\rm i}\; b^\dagger_2
\right)  .
\end{eqnarray}
In terms of these two modes the signal state become therefore
\begin{eqnarray}
\ket{\Psi_0} & = &
\ket{\frac{\alpha}{\sqrt{2}}}\ket{\frac{\alpha}{\sqrt{2}}},\\
\ket{\Psi_1} & = &
\ket{\frac{\alpha}{\sqrt{2}}}\ket{{\rm i}\frac{\alpha}{\sqrt{2}}},\\
\ket{\Psi_2} & = &
\ket{\frac{\alpha}{\sqrt{2}}}\ket{-\frac{\alpha}{\sqrt{2}}},\\
\ket{\Psi_3} & = &
\ket{\frac{\alpha}{\sqrt{2}}}\ket{-{\rm i}\frac{\alpha}{\sqrt{2}}}.
\end{eqnarray}
We can calculate the values of the  $c_j$ in terms of the overlaps of
the four states according to the formula \cite{chefles98a}
$$
   |c_j| = \frac{1}{N^2} \sum_{k,\ell} \exp\left[-\frac{2 \pi {\rm i}
   \;j(k-\ell)}{N} \right] \braket{\Psi_k}{\Psi_\ell}
$$
and find as a function of the expected photon number $\mu = |\alpha|^2$  
\begin{eqnarray}
|c_0| & = & \frac{1}{\sqrt{2}} e^{-\frac{\mu}{4}} \sqrt{\cosh \frac{\mu}{2} +
\cos \frac{\mu}{2}},\\
|c_1| & = &\frac{1}{\sqrt{2}} e^{-\frac{\mu}{4}} \sqrt{\sinh \frac{\mu}{2} +
\sin \frac{\mu}{2}},\\
|c_2| & = &\frac{1}{\sqrt{2}} e^{-\frac{\mu}{4}}  \sqrt{\cosh \frac{\mu}{2} -
\cos \frac{\mu}{2}},\\
|c_3| & = &\frac{1}{\sqrt{2}} e^{-\frac{\mu}{4}} \sqrt{\sinh \frac{\mu}{2} -
\sin \frac{\mu}{2}}.
\end{eqnarray}
The minimum of these four functions depends on the value of
$\mu$. The four functions $4 |c_k|^2$ are plotted in Fig.~\ref{f-1}
from where we can read off $P_D$ as the minimum.

It turns out, however, that for realistic sources these states are not
the correct description of the situation. The reason is that Eve does
not have a phase reference, that means that for a given polarization
she does not see the coherent state $\ket{\alpha}$ but the phase
averaged density matrix $\frac{1}{2 \pi} \int_\phi \ketbra{e^{{\rm i} \phi}
\alpha}{e^{{\rm i} \phi} \alpha }\; {\rm d}\phi$. This results in signal
states which are mixtures of Fock states with a Poissonian photon
number distribution described by the density matrix
\begin{equation}
\rho = e^{- \mu} \sum_n \frac{\mu^n}{n!} \ket{n}\bra{n}.
\end{equation}
Here the state $\ket{n}$ denotes the Fock state with $n$ photons in
one of the four BB84 polarization states. The optimal strategy to
discriminate between the four possible density matrices can be
logically decomposed into  a
QND measurement on the total photon number in the modes $b_1$ and
$b_2$ together and a following measurement which unambiguously discriminates
between the four resulting conditional states for each total photon
number. The justification for this is that the total photon number via
the QND measurement ``comes free'', since the execution of this
measurement does not change the signal states. However, given the
resulting information, we know the optimal strategy on the conditional
states according to \cite{chefles98a}. Therefore we find that the total
probability of unambiguous state discrimination $P_D$
is given in terms of
the respective probabilities for each photon number subspace
$P_D^{(n)}$ as
\begin{equation}
P_D= \sum_{n=0}^\infty e^{- \mu}\frac{\mu^n}{n!} P_D^{(n)}.
\end{equation}
The conditional states resulting from the QND measurement and
corresponding to $n$ photons in total satisfy again the symmetry
condition which allows to apply the results by Chefles and Barnett. We
find for the four coefficients (as a function of the photon number
$n>0$) the expressions
\begin{eqnarray}
|c_0| &=& \sqrt{\frac{1}{4} + 2^{-(1+n/2)} \cos\left( \frac{\pi}{4}
n \right)},\\
|c_1| &=& \sqrt{\frac{1}{4} + 2^{-(1+n/2)} \sin\left( \frac{\pi}{4}
n \right)},\\
|c_2| &=& \sqrt{\frac{1}{4} - 2^{-(1+n/2)} \cos\left( \frac{\pi}{4}
n \right)},\\
|c_3| &=& \sqrt{\frac{1}{4} - 2^{-(1+n/2)} \sin\left( \frac{\pi}{4}
n \right)}.
\end{eqnarray}
Therefore the maximum probability of unambiguous state discrimination
for fixed value of $n$ is given by
\begin{equation}
P_D^{(n)} = \left\{
\begin{array}{l@{\;\;\;}l}
0 & n \leq 2\\
1-2^{1-n/2} & \mbox{$n$ even}\\
1-2^{(1-n)/2}& \mbox{$n$ odd.}
\end{array}
\right.
\end{equation}
It is possible to sum up the contributions from different photon
numbers from the Poissonian distribution and we obtain the expression
\begin{eqnarray}
 \label{PD}
P_D &=& \sum_{n=0}^\infty e^{-\mu} \frac{\mu^n}{n!} P_D^{(n)}
\nonumber \\
 &=& 1 - e^{- \mu} \left( \sqrt{2} \sinh \frac{\mu}{\sqrt{2}}
+ 2 \cosh \frac{\mu}{\sqrt{2}} -1 \right) .
\end{eqnarray}

\begin{figure}[htb]
\centerline{\psfig{width=8cm,file=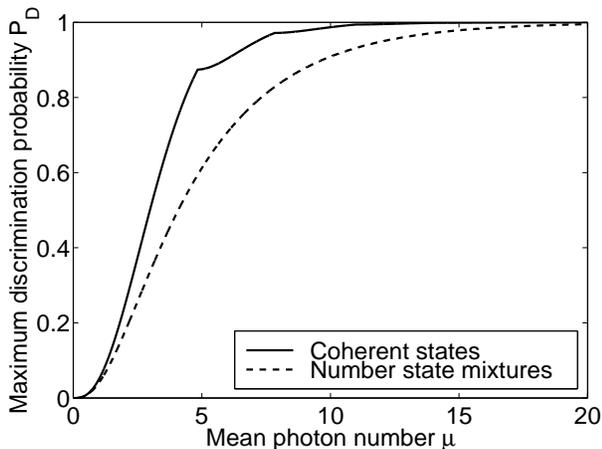}}
\caption{Comparison of the optimum probability of unambiguous states
discrimination for coherent states and for the corresponding mixture
of Fock states. Both have the same Poissonian photon number
distribution with mean photon number $\mu$.}
\label{f0}
\end{figure}

This result is compared to the result for coherent states in
Fig.~\ref{f0}. 
As expected, the probability for unambiguous state identification is
lower for the mixture of Fock-states than for the coherent states. An
expansion  in terms of the photon number $\mu$ gives $P_{D}=\frac{1}{12}
\mu^{3}+ O(\mu^{4})$ for both situations. For lower than third order the
signal states are not linearly independent, so that no unambiguous state
discrimination is possible. Note that an actual implementation does not
necessarily need to follow the decomposition into a QND and another
measurement. We just need to implement one generalized measurement.
Actually, Bennett et al.~\cite{bennett92a} and Yuen \cite{yuen96a} gave
a simple beam-splitter setup which obtains a discrimination probability
of $P_{D}=\frac{1}{32} \mu^{3}+ O(\mu^{4})$.

\section{Unambiguous state discrimination as eavesdropping strategy}
\label{SECstrategy}

We now consider the realistic situation that Alice uses the
phase-averaged coherent states as signal states which are described by a
Poissonian photon number distribution with mean photon number $\mu$. In
this scenario we fix our eavesdropping strategy, to which we refer as
the {\em unambiguous state discrimination attack} (\USD\ attack) as
follows: The unambiguous state discrimination allows Eve to identify a
fraction of the signals without error. For this fraction, she can
prepare a corresponding state close to Bob's detectors such that no
errors appear for these signals. Whenever the identification does not
succeed, she sends the vacuum signal to Bob to avoid errors, which
therefore will not be relevant in the considered scenario.

We need to study this strategy under realistic constraint. An important
constraint is that the transmittance of the quantum channel connecting
both parties is given by the transmission efficiency $\eta_L$. We
consider a detection setup where Bob monitors each polarization mode in
the chosen basis by one detector. These detectors have a finite
detection efficiency $\eta_B$, in which we include any additional loss
on Bob's side, e.g.~from a polarizing beamsplitter. The detectors are
modeled as  ``yes/no'' detectors, which either fire, or they do not
fire; they cannot distinguish the number of impinging photons.

It is clear that once Eve identified a signal she is interested to
produce a signal in the corresponding polarization such that Bob will
detect it despite his inefficient detectors. One strategy is to send a
stronger signal than the original one in the correct polarization. This
will work as long as Bob measures in the polarization basis which
includes the signal polarization (sifted key), but it will lead to an
increased coincidence rate of clicks in both of Bob's detectors
otherwise. Our analysis extends previous analysis to include the
additional constraint put on the eavesdropping strategy by the  fact
that Bob observes not only the rate of clicks of one or the other
detector, but also the rate of events when both detectors fire, each
monitoring one of the orthogonal polarization modes. The latter event
will be observed ideally only when Alice and Bob use different bases,
independently of the presence of absence of an eavesdropper. Eve's aim
is to reproduce these two observables with the minimum number of
non-vacuum signals to make efficient use of the successfully identified
signals.

In the absence of Eve, whenever Alice and Bob use the same polarization
basis Bob's expects to find at most one detector clicks; the probability
of a click is
\begin{equation}
     \bar{P}_1 = 1 - \exp( - \eta_L \eta_B \mu ), \label{P1bar}
\end{equation}
as follows from the Poissonian photon-number statistics of coherent
states.

Whenever Alice and Bob use different bases a double-click may occur; its
probability is
\begin{equation}
     \bar{P}_2 = \left[ 1 - \exp \left( - { \eta_L \eta_B \mu \over 2 } \right)
     \right]^2.
  \label{P2bar}
\end{equation}

What happens in the presence of Eve depends on the signals Eve sends for
the successfully detected Alice's signals. It is clear that Eve can
avoid the occurrence of double clicks when Alice and Bob measure in the
same basis, since she unambiguously determined the signal. Therefore it
is not useful to monitor the double click rate when Alice and Bob use
the same basis.

Note that we do not need to include detector dark count rates or
errors due to misalignment into account. The reason for that is that
we will investigate the limit when the \USD\  attack gives complete
information to Eve while it reproduces the expected  probabilities
$\bar{P}_1$ and $\bar{P}_2$.
The values of these  probabilities in the absence of an eavesdropper and
the reproduced values resulting from a successful \USD\  attack will be
affected in the same way by the error mechanisms of dark counts and
misalignment etc., so that the resulting real observed rates will still
be indistinguishable.

\subsection{Eve sends $n$-photon states}

Let us suppose now, that whenever Eve succeeds in the unambiguous state
discrimination she sends a number state (with correct polarization)
containing $N$ photons to Bob. If she fails she simply sends no photon.

If Alice and Bob use the same basis, at most one of two Bob's
detectors will click.
The probability of this event is given by
\begin{eqnarray}
     P_1^{(N)} &=& P_D \left[ 1 - {m \choose 0} \eta_B^0 (1 -
           \eta_B)^{N} \right] \nonumber \\
         &=& P_D \left[ 1 - (1-\eta_B)^N \right]  .
 \label{P1}
 \end{eqnarray}
This is the probability that one detector clicks if a state
$|N\rangle$ comes, multiplied by the probability that Eve succeeds in \USD\
(and sends $|N\rangle$).

If Alice and Bob use different bases, we can think of the photons as
being  equally and   independently
distributed  to both Bob's detectors. The probability to find $k$ photons
at the first detector and $\ell$ photons at the second one (with included
detection efficiencies)  is given by the formula
 \begin{eqnarray}
   \Pi_{k\ell} =&& P_D \left( \frac{1}{2} \right)^{\! N}
                \sum_{m=k}^{N-\ell} {N \choose m}
                 \left[
                {m \choose k} \eta_B^k (1-\eta_B)^{m-k}
                 \right]
                 \times \nonumber \\
                && \left[
                {N-m \choose \ell} \eta_B^\ell (1-\eta_B)^{N-m-\ell}
                 \right]. \nonumber
 \end{eqnarray}
where the summation limits stem from obvious constraints $m \ge k, N-m
\ge \ell$.
Thus the probability of double click in Bob's ``yes-no'' detectors
when Eve is active and while Alice  and Bob use different
polarization bases reads
 $$
     P_2^{(N)} = P_D \left[ 1 - \sum_{\ell=0}^{N} \Pi_{0\ell} -
          \sum_{k=0}^{N} \Pi_{k0} + \Pi_{00} \right],
 $$
(note that $\Pi_{00}$ would be subtracted two times).
Because of the symmetry of the configuration, obviously, $\sum_{\ell=0}^{N}
\Pi_{0\ell} = \sum_{k=0}^{N} \Pi_{k0}$.  With the expressions
 $$
    \Pi_{00} = P_D (1 - \eta_B)^N,
 $$
 \begin{eqnarray}
    \sum_{k=0}^{N} \Pi_{k0} &=& P_D \sum_{m=0}^{N} {N \choose m} 2^{-N}
           \sum_{k=0}^{m} {m \choose k}
           \eta_B^{k} (1 - \eta_B)^{N-k} \nonumber \\
           &=& P_D
           \left( 1 - {\eta_B \over 2} \right)^N \nonumber
 \end{eqnarray}
we obtain finally for the   double-click probability
  \begin{equation}
     P_2^{(N)} = P_D \left[ 1 - 2\left( 1 - {\eta_B \over 2} \right)^N
          + (1-\eta_B)^N \right] .
  \label{P2}
  \end{equation}

\subsection{Eve sends a mixture of number states}

Of course, there is no reason to restrict Eve only to the use of number
states. After successful state discrimination she can send to Bob any
pure state or mixture. However, from Bob's point of view these signals
are effectively mixtures of photon number states because of the nature
of his detectors (they may be described by the pair of projectors:
${\sf P}_{\rm no}=| 0 \rangle \langle 0 |$ and
${\sf P}_{\rm yes}=\sum_{n=1}^\infty | n \rangle \langle n |$).
Therefore it is sufficient to  analyze only a mixture of photon number
states in the polarization of the identified signal, so that only the
photon number statistics remains to be  chosen by Eve. 

\begin{figure}[t] 
\centerline{\psfig{width=8cm,file=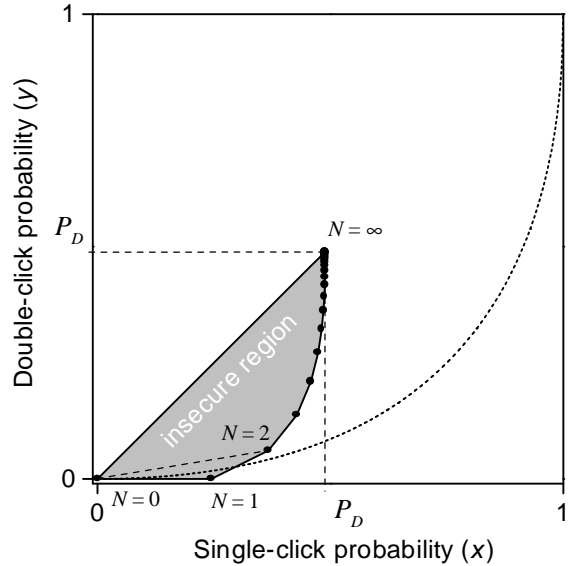}}
\caption{Diagram displaying relations between ``single-click'' and
``double-click'' probabilities. The highlighted area contains all
possible combinations of Bob's detection probabilities stemming from
Eve's activity (described in the text) for a given detection efficiency
(here, particularly, $\eta_B=0.5$) and a given mean photon number in
states sent by Alice ($\mu=4$). It is a region of {\em insecure} key
generation. The shape of the area depends on $\eta_B$, the scaling on
$\mu$ [through discrimination probability $P_D(\mu)$]. The separate
dotted curve represents a set of all possible ``working points'' without
an eavesdropper, i.e.\ a set of all possible pairs of expected
$\bar{P}_1$ and $\bar{P}_2$. Any particular position of a working point
depends on the values of the line transmittance ($\eta_L$), the
detection efficiency ($\eta_B$), and the mean photon number ($\mu$). The
value of $\mu=4$ is chosen to make the diagram well readable. The
structure is the same for lower, realistic values.}
\label{f2}
\end{figure}

As already mentioned, Bob is interested only in the number of single
clicks (in case that his and Alice's bases coincide) and double clicks
(if the bases differ). One can plot very illustrative diagram displaying
relations between corresponding single-click and double-click
probabilities (see Fig.~\ref{f2}).

The situation where Eve sends number states to Bob is represented by a
dot for each value of the photon number $N$. The positions of these dots
have been calculated for fixed values of $\eta_L$ and $\mu$. Coordinates
of a point corresponding to any mixture of number states can always be
expressed as a linear convex combination of coordinates corresponding to
individual number states. Because of the convexity of the above
mentioned curve all such points must lie inside (or on the boundary) of
the polygon with vertices at the points corresponding to number states
(i.e.\ in the area highlighted in Fig.~\ref{f2}).

We can explicitly prove the convexity of the boundary formed by the
points for fixed photon number. The points with $x$-coordinate
$P_1^{(N)}$ [Eq.~(\ref{P1})] and $y$-coordinate $P_2^{(N)}$
[Eq.~(\ref{P2})] lie on a continuous curve which can be expressed with
help of  Eq.~(\ref{P1}) by a real continuation of the parameter $N$ as
 $$
    N = \frac{\ln(1-x/P_D)}{\ln(1-\eta_B)}.
 $$
Substituting into Eq.~(\ref{P2}) we obtain the explicit
equation of the curve
  \begin{equation}
      y = \left[ 2 - 2 \left( 1 - {x \over P_D}  \right)^\kappa -
          {x \over P_D}  \right] P_D,
   \label{N-curve}
  \end{equation}
where
 $$
    \kappa = \frac{\ln(1-\eta_B/2)}{\ln(1-\eta_B)}.
 $$
Calculating the second derivative of Eq.~(\ref{N-curve}) with respect to
$x$ and using the fact that $\eta_L$, $\eta_B$, and $x/P_D$ take values
in the interval between 0 and 1, it follows that the curve given by
Eq.~(\ref{N-curve}) is convex. This proves that the highlighted area in
Fig.~\ref{f2} is indeed convex.

\subsection{Insecure parameter regime}

The convex area defined in the previous section can be called a region
of insecurity. We define the  working point of a setup as the  the point
whose coordinates are given by expected values in the absence of an
eavesdropper. If this working point falls into the region of insecurity,
Eve can get complete information on the key without a risk of being
disclosed.

The set of all possible working points is represented by the dotted
curve in the diagram. Expected single-click probability $\bar{P}_1$
[Eq.~(\ref{P1})] represents $x$-coordinate, expected double-click
probability $\bar{P}_2$ [Eq.~(\ref{P2})] represents $y$-coordinate. From
Eq.~(\ref{P1bar}) the exponential can be expressed and substituted into
Eq.~(\ref{P2bar}). Thus the explicit equation of the working point curve
reads
 \begin{equation}
      y = \left[ 1 - (1 - x)^{1/2} \right]^2.
  \label{W-curve}
 \end{equation}

We have to answer the question  for which values of parameters $\eta_L$, $\eta_B$,
and $\mu$ the working point lies in the region of insecurity.

\subsubsection{Necessary  condition for insecurity}

\begin{figure}[t] 
\centerline{\psfig{width=8cm,file=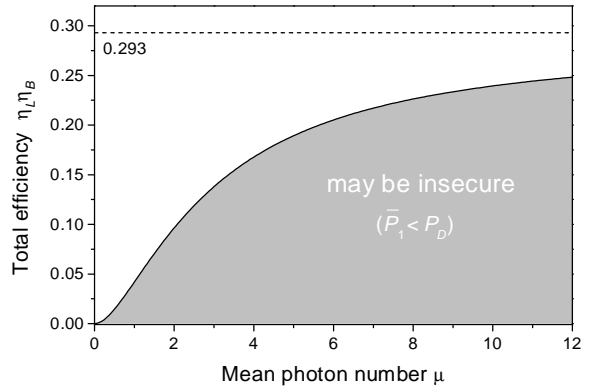}}
\caption{If the value of expected single-click probability is greater
    then the discrimination probability ($\bar{P}_1 > P_D$) the
    described \USD\ attack can be, in principle, detected. The plot shows
    an example of a curve separating the set of values of total
    efficiencies ($\eta_L \eta_B$) and mean photon numbers (in states sent by
    Alice) satisfying the above constraint [see inequality
    (\protect\ref{raw})].}
\label{f3}
\end{figure}

If the expected probability of single clicks  satisfies $\bar{P}_1 >
P_1^{(N)}$ for all $N$ then the working point will certainly not fall to
the region of insecurity, which is clearly illustrated in Fig.~\ref{f2}.
This leads  to the necessary condition for insecurity given by
$\bar{P}_1 < P_D$. To evaluate the implication for the experimental
parameters, we substitute Eq.~(\ref{P1bar})
 \begin{equation}
        \eta_L \eta_B < \frac{-\ln[ 1 - P_D(\mu) ]}{\mu}.
  \label{raw}
 \end{equation}
An analysis of this expression shows 
that for a fixed expected photon number $\mu$ a system cannot be
cracked by an \USD\  attack if the total transmission efficiency $\eta_L \eta_B$
is higher than a certain threshold which depends on the the expected photon
number $\mu$. This dependence is evaluated numerically in Fig.~\ref{f3}.
The surprising aspect is, that the threshold does not
go to $1$ as $\mu$ goes to infinity. Instead we find 
 \begin{eqnarray}
     (\eta_L \eta_B)^{(\infty)} &=& \lim_{\mu \to \infty}
         \frac{-\ln[ 1 - P_D(\mu) ]}{\mu} \nonumber \\
         &=& (1 - 2^{-1/2})
         \approx 0.293  .
   \label{lim_losses}
 \end{eqnarray}
This shows, that that the implementation of quantum cryptography with
weak coherent states cannot be cracked completely by the \USD\  attack for
{\em all} values of the expected photon number $\mu$ as  long as the
total transmission satisfies $\eta_L \eta_B\geq 1 - 2^{-1/2}$.

\subsubsection{Sufficient condition of insecurity}

In this section we will derive precise conditions determining
when a working point falls into the region of insecurity. In a first
step we will show that for parameters of practical applications it is
sufficient to consider the scenario  the working point falls below the
straight line going through the origin and the vertex $N=2$.
This condition corresponds to
 \begin{equation}
    x_w \ge y_w \frac{x_2}{y_2}.
  \label{cond1}
 \end{equation}
The coordinates of points used in this condition are defined in Tab.~\ref{tab}.
In the second step we can then determine whether
in this scenario the working point
lies inside or outside the region of insecurity by
checking on which side of the line going through the vertices $N=1$ and
$N=2$ it lies (see Fig~\ref{f2}).
If it lies on the left, \QKD\ is insecure.
This corresponds to the inequality
\begin{equation}
    x_w \le y_w \frac{x_2 - x_1}{y_2} + x_1.
  \label{cond2}
\end{equation}

\begin{table}
  \begin{tabular}{lll}
    Working point: & $x_w =$ & $y_w =$ \\
          & $~~1 - \exp( - \eta_L \eta_B \mu )$ &
         $~~\left[ 1 - \exp \left( - { \eta_L \eta_B \mu \over 2 } \right)
         \right]^2$ \\
    Vertex $N=1$: & $x_1 = P_D \eta_B$ & $y_1 = 0$ \\
    Vertex $N=2$: & $x_2 = P_D (2\eta_B - \eta_B^2)$
            &  $y_2 = P_D \eta_B^2 / 2$ \\
  \end{tabular}

   \caption{Coordinates of selected  points in the parameter space of 
   ``observables'', which are the probabilities of
   single clicks ($x$) and double clicks ($y$) in Bob's detectors.}
   \label{tab}
 \end{table}

First, let us turn to the  inequality (\ref{cond1}). Substituting expressions
for all coordinates according to Tab.~\ref{tab} one obtains
an inequality which is quadratic in the variable  
  $  R = \exp \left( -{\eta_L \eta_B \mu / 2} \right)$ with the
parameter $\eta_B$. We find that the working point lies below the line
connecting vertices $N=0$ and $N=2$ if
$R \in \left( \frac{4-3\eta_B}{4-\eta_B}, 1  \right)$.
Thus the mean photon number in coherent states
sent by Alice must be lower than a threshold $\mu_2$ given by
 \begin{equation}
\label{mucond}
       \mu < \mu_2 = \frac{-2}{\eta_L \eta_B} \ln \left(
       \frac{4-3\eta_B}{4-\eta_B} \right)  .
  \label{mu2}
 \end{equation}
We find that $\mu_2 \in [1/\eta_L, 2 \ln 3 / \eta_L]$ for any $\eta_B$
and, especially, always $\mu_2 \geq 1$.
As we can see, this condition is
satisfied in all current experiments and does not pose a serious
restriction to the validity of our analysis especially for non
negligible loss.

Now let us turn our attention to the condition (\ref{cond2}) which,
whenever  condition (\ref{mucond}) is fulfilled, determines whether
the working point is in the region of insecurity. It can be
expressed in the following form
 \begin{equation}
\label{Fcond}
      F(\mu,\eta_L,\eta_B) := x_w \eta_B - 2 y_w (1-\eta_B) - P_D
      \eta_B^2 \le 0,
   \label{F}
 \end{equation}
Due to the complicated dependence of $P_D$ on $\mu$ we failed to find
its analytical solution. The analytical statement we can do without
      any extra approximation is based on the observation that
$$
   \left. \frac{\partial F}{\partial \mu} \right|_{\mu=0} \! = \eta_L
      \eta_B^2 > 0
   \quad {\rm and} \quad
   F(0,\eta_L,\eta_B) = 0  .
 $$
This implies that there exists always a range of values for $\mu$
starting from $\mu = 0$ for
which we have $F>0$, i.e.\ the security of the key distribution cannot
be cracked completely by the \USD\ attack.

\begin{figure}[t] 
\centerline{\psfig{width=8cm,file=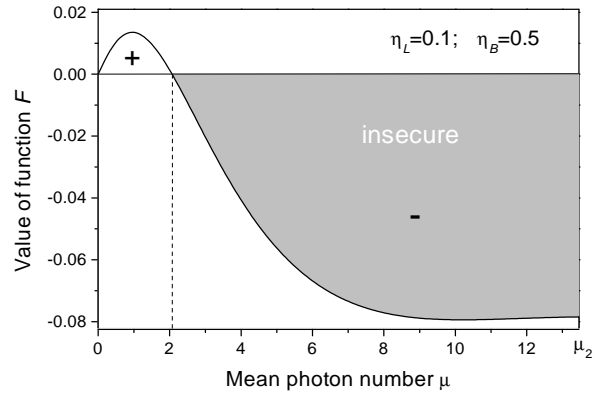}}
\caption{The sign  of the function $F(\mu,\eta_L,\eta_B)$ is a
criterion for the security (positive) or insecurity (negative) of the
the quantum key distribution with respect to the \USD\  attack. The line
   transmittance and Bob's detection efficiency are fixed: $\eta_L=0.1$,
   $\eta_B=0.5$. Mean photon number, $\mu$, goes from zero to $\mu_2$
   limit. If $F$ is negative the transmission is totally
   insecure. The zero point lies at $\mu \approx 2.07$ photons.}
\label{f5}
\end{figure}

It is easy to evaluate condition (\ref{Fcond})  numerically. In
Fig.~\ref{f5} we give  an example for the values of line transmittance
$\eta_L = 0.1$ and detection efficiency $\eta_B = 0.5$ (so that  $\mu_2
\approx 13.46$). In this particular case, the transmission becomes
insecure in about 2.07 photons.
It is not completely satisfying to have to fall back to numerical
methods to investigate the security against the \USD\
attack. Fortunately, it is possible to get some analytic results in a
situation which is relevant to applications.

\subsubsection{Partly accessible loss in a system with large loss}

The results of the previous sections illuminate to which extend Eve can
achieve perfect eavesdropping by making an   unambiguous state
discrimination measurement followed by  sending the identified signals
directly to Bob's detectors, thereby bypassing the lossy quantum
channel.

However, Eve does not necessarily need to access the whole lossy quantum
channel to be successful. By accessing we mean, that Eve can avoid these
losses either by replacing a quantum channel by a perfect, loss-free
one, or by replacing it by classical communication and state
preparation. The formulas of the previous sections still apply if we
collect in the quantity $\eta_B$ all those losses on the way to Bob's
detector which are not accessible to Eve, while $\eta_L$ denotes now
only that loss that is accessible to her. It is instructive to look at
the limit of high non-accessible losses ($\eta_B \ll 1$). In that case
we can approximate the function $F$ of equation (\ref{Fcond}) by
\begin{equation}
 F \approx \eta_B^2 \left(\eta_L \mu - \frac{1}{2} \eta_L^2 \mu^2 -
 P_D\right).
\end{equation}
The insecurity criterion $F \le 0$ in the region $\mu<\mu_2$ (from
Eqn.~(\ref{mucond})) then leads to the condition
\begin{equation}
\label{etacond}
\eta_L \le \frac{1}{\mu}\left( 1 - \sqrt{1- 2 P_D}\right) ,
\end{equation}
which is independent of $\eta_B$. It can be approximated by
\begin{equation}
\eta_L \le \eta_L^{\rm crit} \approx \frac{P_D}{\mu} \approx
\frac{1}{12} \mu^2 .
\end{equation}

\begin{figure}[htb]
\centerline{\psfig{width=8cm,file=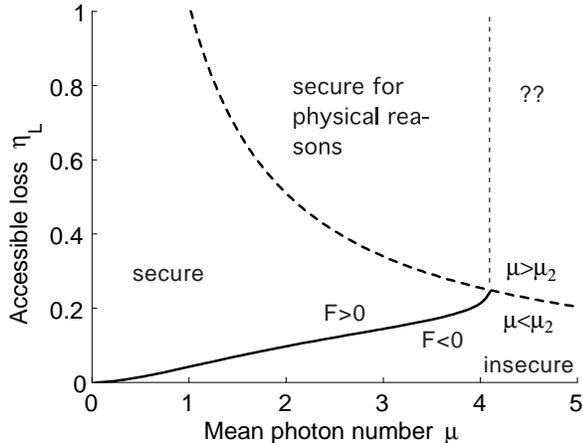}}
\caption{The secure parameter regime for the losses accessible to Eve
for large Bob's losses ($\eta_B \ll 1$) is the region above the solid line
($F>0$). In the region with $F \le 0$ and $\mu < \mu_2$ the system is
insecure. In the remaining region we have $F\le 0$, but since
$\mu>\mu_2$, we cannot make any definitive statements about security. }
\label{USDadd}
\end{figure}

Condition (\ref{etacond}) is shown in Fig.~\ref{USDadd} as a solid
line.  To make statements about security against the \USD\ attack, we
need to consider additionally condition (\ref{mucond}), which can be
approximated by $\mu < \frac{1}{\eta_L}$ in leading order of $\eta_B$
and is shown as a dashed line.  We now can conclude that the system is
secure against \USD\ attacks in the regime of small detection
efficiencies $\eta_B$ if we are in the parameter region with $F>0$ and
$\mu<\mu_2$. Furthermore, the system is insecure in the region $F \le 0$
and $\mu<\mu_2$. For the region with $\mu>\mu_2$ we can only make
indirect statements. One is, that if the system is secure for a pair
of values $(\mu, \eta_L)$, then it must be secure for all values
$(\mu, \eta_L')$ with $\eta_L' > \eta_L$, otherwise Eve could gain an
advantage by not accessing all the loss available to her. Therefore
the only region about which we cannot make a statement with the
present calculation is the region with $\mu > 4.1$ and
$\eta_L>\frac{1}{\mu}$. Here more detailed calculation would be necessary.

Note that these considerations are valid for $\eta_B \ll 1$ and only in
this limit $\eta_B$ does no longer play any role. For higher values of
$\eta_B$ this changes.

\subsection{Remark to the statistical nature of the problem}

One should keep in mind that all of Bob's measurements have a
statistical character. Bob does not measure probabilities but finite
numbers of clicks which naturally fluctuates. In practice Bob must set
certain limits of a ``confidential interval'' of acceptable numbers of
detector clicks. This effects that in some cases Bob will reject the
transmission even if no eavesdropper is present. A more serious
implication  is that there is always some non-zero probability that Eve
will not be detected even if the working point lies outside the
insecurity region.

Note that  Eve does not need to eavesdrop all the time -- she may let
pass a fraction of the signal sequence without any intervention. Her
(deterministic) information on the key decreases with this strategy. But
both Bob's single- and double-click probabilities also change. The point
corresponding to such an eavesdropping strategy (in the diagram as in
Fig.~\ref{f2}) shifts along  the straight line connecting ``full-time''
Eve's strategy point with Alice's and Bob's working point. The relative
shift equals the fraction of transmission during which Eve is active.

For practical purposes it would be necessary to determine the
probability that Eve's information on the key (due to the \USD\ attack)
will be smaller than a certain chosen limit, as a function of the limits
of the confidential interval and of the length of the key. This
represents a challenge for the further research in this field.

\section{USD attack versus beamsplitting attack}
\label{SECUSD}

Traditionally, security against the beamsplitting attack
\cite{bennett92a} has been used as a practical level of security. In the
beamsplitting attack the lossy line is replaced by an ideal loss-free
line complemented by a beamsplitter such that the total loss of the
original line is reproduced. The eavesdropper stores any photons coming
out of the free arm of the beamsplitter. Whenever the eavesdropper {\em
and} the receiver obtain a photon, which is possible for multi-photon
signals, Eve can measure her signal after she learns the polarization
basis in the public announcement and  she will learn thereby the bit
value of these signals completely.

It is interesting to note that security against beamsplitting attack
suggests that one can obtain a secure key even for large average
photon numbers. In the absence of errors, the gain rate of secure key
bits per signal bit can be approximated in a way similar to that used
in \cite{nlsub00a} for the optimal individual attack. This
approximation is given by
\begin{equation}
G_{\rm BS} = \frac{1}{2} \left(p_{\rm exp} - p_{\rm split}\right),
\end{equation}
where the factor $1/2$ stems from discarding signals with unequal
polarization basis. Then $p_{\rm exp}$ is the probability that Bob
receives a signal, while $p_{\rm split}$ is the joint probability that
Eve learned the bit value of a signal and that the signal is received by
Bob.  To point out the basic problem of the beamsplitting attack it is
sufficient to consider the case of $\eta_B =1$ and of coherent states.
Then we find for Poissonian photon statistics and a transmission rate
$\eta$ of the system
\begin{eqnarray}
p_{\rm exp}& = & 1-\exp(-\eta \mu),\\
p_{\rm split}& = &p_{\rm exp}\left\{1-\exp\left[-(1-\eta) \mu
\right]\right\},\\
G_{\rm BS}& =& \frac{1}{2}\exp(-(1-\eta) \mu) \left[1-\exp(-\eta \mu)
\right]  ,
\end{eqnarray}
which is always positive. Actually, the optimum is obtained for $\mu
\approx 1$. It is clear from our analysis, however, that for large
values of $\mu$ and typical loss rates, the \USD\  attack will render the
quantum key distribution protocol completely insecure.

The awareness of this problem is low, and it is thought that it can be
avoided by complementing the beamsplitting attack with the additional
requirement to keep the average photon number low, much lower than $1$,
as to keep the set-up in the quantum domain. This seems  rather odd,
since there is no obvious justification for this requirement. More
importantly, even for photon numbers $\mu \ll 1$ we find that for
sufficiently large loss the transmission becomes insecure according to
the \USD\  attack while the analysis according to the beamsplitting attack
makes us believe that we are dealing with a secure key. It seems that
the \USD\  attack is underestimated since the probability of success in the
unambiguous state discrimination goes with $\mu^3$ since only for three
or more photons the four signal states are actually linear independent.
The beamsplitting attack, however, succeeds with a probability of order
$\mu^2$, since already two photons can be split by the beamsplitter.

This seems to imply that beamsplitting is the more powerful attack.
However, this is not the case since the two attacks vary in their power
differently as the loss of the system increases. In the \USD\  attack the
probability to identify  a signal depends only on the average photon
number $\mu$, and once this probability is high enough to generate the
expected number of signals for the receiver (which depends on the amount
of loss) then the transmission becomes insecure. In the beamsplitting
attack, on the other hand, the total probability of identified signals
$p_{\rm split}$ depend on $\mu$ {\em and}  on the transmission
coefficient $\eta$, and this probability goes {\em down} with increasing
loss for fixed $\mu$. And indeed, we find that $p_{\rm exp} > p_{\rm
split}$. In other words,  the beamsplitting attack becomes less
efficient with increasing loss. This is easy to see in a simple example
of a two-photon signal. The probabilities $p(n,2-n)$ that $n=0,1,2$
photons arrive at Bob's detectors and $n-2$ photons go to Eve in the
beamsplitting attack are given by
\begin{eqnarray}
p(0,2) & = & (1-\eta)^2,\\
p(1,1) & = & 2 \eta (1-\eta), \\
p(2,0) & = &  \eta^2.
\end{eqnarray}
This means, that for high losses ($\eta \ll1$)
most likely both photons are sent to Eve.
Probability of this event is $p(0,2) \approx 1-2 \eta$,
while the splitting probability goes down as
$p(1,1) \approx  2 \eta$. The respective probabilities for $n$-photon
signals are of the same order of magnitude in $\eta$. 
Therefore, clearly, there is a crossover as a function of $\eta$ where
for fixed average photon number $\eta$ the \USD\  attack is more
efficient than the beamsplitting attack.

We would like to stress again that from a technological point of view
the \USD\  attack seems to be easier to implement than the beamsplitting
attack. This is based on two points. Firstly, experience indicates that
a complete measurements destroying the quantum state completely, as
possible by the \USD\  attack,  are easier to realize (at least in some
approximation) than the realization of a quantum channel with reduced
loss, as required by the beamsplitting attack. Secondly, the
beamsplitting attack implies the use of quantum memory, which could
store the split-off signal photons until the polarization bases for each
signal are announced.

Finally we would like to point out again that a security proof for
realistic signals with Poissonian photon number distribution
exists for individual attacks \cite{nlsub00a} and coherent attacks
\cite{inamorisub00a}. Naturally, these security proofs include the
security against the beamsplitting attack and against the  \USD\  attack.

\section{Conclusions}
\label{SECconclusions}

We have quantitatively analyzed an attack against realistic quantum
crypto systems which enables an eavesdropper to gain information on the
key without causing any errors in case of a lossy channel or poor
detection efficiencies. It uses unambiguous discrimination of linearly
independent signal states. This attack does not require the ability to
store  quantum states or to perform complicated quantum dynamics.
Moreover, the attack does not require  to substitute the lossy quantum
channel by a perfect one.

We have derived a set of conditions which allow to judge whether a given
system can be totally insecure under the \USD\  attack. We have shown a
secure parameter regime in terms of the total transmission efficiency
and the mean photon number. In the important limit of small detection
efficiencies $\eta_B$ we have obtained an analytic result so that we can
give  explicitly a set of parameters (line transmittances, detector
efficiencies and mean photon numbers in coherent states sent by Alice)
for which the transmission is secure/insecure under the \USD\  attack. In
theory, the signal can always be chosen to be weak enough to allow
secure communication. In practice, however, the detector noise places
restrictions on that end \cite{brassardsub99a}. Finally, we showed that
security against beamsplitting attacks does not necessarily imply
security against the \USD\  attack. This implies that we need to search for
a better conditional security criterion against attacks deemed
practically with currently available technology.

\section*{Acknowledgments}

M.D.\ acknowledges discussions with Ond\v{r}ej Haderka and Martin
Hendrych and support from grants of the Czech Ministry of Education
(research project ``Wave and particle optics'' and project VS 96028) and
the Czech National Security Authority (19982003012). This work has been
supported under project 43336 by the Academy of Finland.




\begin{thebibliography}{10}

\bibitem{vernam26a}
G.~S. Vernam, Journal of the American Institute of Electrical Engineers {\bf
  45},  109  (1926).

\bibitem{bennett84a}
C.~H. Bennett and G. Brassard,  in {\em Proceedings of IEEE International
  Conference on Computers, Systems, and Signal Processing, Bangalore, India}
  (IEEE, New York, 1984), pp.\ 175--179.

\bibitem{wiesner83a}
S. Wiesner, Sigact News {\bf 15},  78  (1983).

\bibitem{huttner94a}
B. Huttner and A.~K. Ekert, J. Mod. Opt. {\bf 41},  2455  (1994).

\bibitem{mayers96a}
D. Mayers,  in {\em Advances in Cryptology \,---\, Proceedings of Crypto '96}
  (Springer, Berlin, 1996), pp.\ 343--357, available as quant-ph/9606003.

\bibitem{mayers98a}
D. Mayers, Unconditional security in Quantum Cryptography, Report
  quant-ph/9802025v4, (1998).

\bibitem{lo99a}
H.-K. Lo and H.~F. Chau, Science {\bf 283},  2050  (1999).

\bibitem{huttner95a}
B. Huttner, N. Imoto, N. Gisin, and T. Mor, Phys. Rev. A {\bf 51},  1863
  (1995).

\bibitem{yuen96a}
H.~P. Yuen, Quantum Semiclassic. Opt. {\bf 8},  939  (1996).

\bibitem{dusek99a}
M. Du\v{s}ek, O. Haderka, and M. Hendrych, Opt. Comm. {\bf 169},  103  (1999).

\bibitem{brassardsub99a}
G. Brassard, N. L\"utkenhaus, T. Mor, and B. Sanders, Security Aspects of
  Practical Signal Sources for Quantum Cryptography, In preparation, 1999.

\bibitem{nlsub00a}
N. L\"utkenhaus, Security against individual attacks for realistic quantum key
  distribution, To appear Phys.~Rev.~A. Also quant-ph/9910093.

\bibitem{inamorisub00a}
H. Inamori, N. L\"utkenhaus, and D. Mayers, In preparation.

\bibitem{bennett92a}
C.~H. Bennett, F. Bessette, G. Brassard, and L. Savail, J. Cryptology {\bf 5},
  3  (1992).

\bibitem{ivanovic87a}
I.~D. Ivanovic, Phys. Lett. A {\bf 123},  257  (1987).

\bibitem{peres88a}
A. Peres, Phys. Lett. A {\bf 128},  19  (1988).

\bibitem{jaeger95a}
G. Jaeger and A. Shimony, Phys. Lett. A {\bf 197},  83  (1995).

\bibitem{chefles98a}
A. Chefles and S.~M. Barnett, Phys. Lett. A {\bf 250},  223  (1998).

\bibitem{huttner96a}
B. Huttner, A. Muller, J.~D. Gautier, H. Zbinden, and N. Gisin, Phys. Rev. A
  {\bf 54},  3783  (1996).

\bibitem{brandt99a}
H.~E. Brandt, Am. J. Phys {\bf 67},  434  (1999).



\end{thebibliography}
\end{document}